\documentclass{emulateapj}

\begin{document}

\shortauthors{Luhman, Burgasser, \& Bochanski}
\shorttitle{Candidate for Coolest Brown Dwarf}

\title{Discovery of a Candidate for the Coolest Known Brown
Dwarf\altaffilmark{1}}

\author{
K. L. Luhman\altaffilmark{2,3},
A. J. Burgasser\altaffilmark{4,5},
and J. J. Bochanski\altaffilmark{2}
}

\altaffiltext{1}{Based on observations made with the Spitzer Space
Telescope, which is operated by the Jet Propulsion Laboratory,
California Institute of Technology under a contract with NASA, and
with the 6.5 meter Magellan Telescopes located at Las Campanas Observatory,
Chile.}
\altaffiltext{2}{Department of Astronomy and Astrophysics, The Pennsylvania
State University, University Park, PA 16802, USA; kluhman@astro.psu.edu}
\altaffiltext{3}{Center for Exoplanets and Habitable Worlds, The 
Pennsylvania State University, University Park, PA 16802, USA}
\altaffiltext{4}{Center for Astrophysics and Space Science, University of
California San Diego, La Jolla, CA 92093, USA}
\altaffiltext{5}{Massachusetts Institute of Technology, Kavli Institute for
Astrophysics and Space Research, 77 Massachusetts Avenue, Cambridge, MA 02139,
USA}

\begin{abstract}

We have used multi-epoch images from the Infrared Array Camera on board
the {\it Spitzer Space Telescope} to search for substellar companions
to stars in the solar neighborhood based on common proper motions.
Through this work, we have discovered a faint companion to the
white dwarf WD~0806-661. The comoving source has a projected separation of
$130\arcsec$, corresponding to 2500~AU at the distance of the primary (19.2~pc).
If it is physically associated, then its absolute magnitude at 4.5~\micron\ is
$\sim1$~mag fainter than the faintest known T dwarfs, making it a
strong candidate for the coolest known brown dwarf. The combination of
$M_{4.5}$ and the age of the primary (1.5~Gyr) implies an effective temperature
of $\sim300$~K and a mass of $\sim7$~$M_{\rm Jup}$ according to theoretical
evolutionary models.
The white dwarf's progenitor likely had a mass of $\sim2$~$M_\odot$,
and thus could have been born with a circumstellar disk that was sufficiently
massive to produce a companion with this mass. Therefore, the 
companion could be either a brown dwarf that formed like a binary star or
a giant planet that was born within a disk and has been dynamically
scattered to a larger orbit.

\end{abstract}


\section{Introduction}
\label{sec:intro}

The methane-bearing T dwarfs comprise the coolest known class of brown dwarfs
\citep[][references therein]{bur06}. These sources have atmospheres that
are sufficiently cool for H$_2$O and CH$_4$ gases to form in abundance,
resulting in distinct, planet-like spectra \citep{opp95,geb96}.
The similar masses and effective temperatures of T dwarfs as
compared to ``warm" exoplanets make them critical benchmarks for theoretical
atmospheric and interior models of substellar objects \citep{mar96,all96}.
Although more than 200 T dwarfs have been discovered over the last decade,
there remains a large gap in temperature between the coolest known T dwarfs
\citep[$T_{\rm eff}\sim500$~K][]{burn08,del08,luc10}
and the Jovian planets ($T_{\rm eff}\sim150$~K).
Based on theoretical spectra, objects in this temperature range may exhibit
spectroscopic characteristics that are distinct from those of T dwarfs,
including strong near-IR NH$_3$ absorption and scattering from water ice clouds
\citep{bur03}. These sources may therefore represent a
new spectral class tentatively designated as Y dwarfs \citep{kir99,kir05}.
In addition, the coolest brown dwarfs encompass the oldest and lowest mass
``stars" in the Galaxy, probing the history and efficiency of star formation at
the planetary-mass limit.

Wide-field imaging surveys over the last decade have been successful in
uncovering free-floating brown dwarfs at progressively cooler temperatures.
However, the relatively low surface density of field brown dwarfs on
the sky \citep{met08} necessitates imaging of very large areas and spectroscopy
of samples of candidates that can be highly contaminated.
A survey for companions to stars in the solar neighborhood ($d\lesssim30$~pc)
is an appealing alternative for identifying the coolest brown dwarfs because
of the smaller search volume that is required.
In addition, the age, distance, and metallicity measured for a primary
can be adopted for its companion, parameters which are usually
difficult or impossible to measure for isolated brown dwarfs.
Several of these benchmark T dwarfs, including the prototype Gliese 229B,
have been discovered as companions to nearby main sequence stars 
\citep{nak95,opp95,wil01,bur00,bur05,bur10,sch03,mcc04,bil06,mug06,luh07,burn09,gol10,sch10}
as well as a white dwarf \citep{day11}.

The Infrared Array Camera  \citep[IRAC;][]{faz04} on board the
{\it Spitzer Space Telescope} \citep{wer04} offers the best available
sensitivity to cool companions in wide orbits ($T_{\rm eff}<1000$~K,
$r>100$~AU). We have used multi-epoch images from IRAC to search for
substellar companions in the solar neighborhood based on common proper motions.
In this Letter, we present the discovery of a new companion
that is a candidate for the coolest and faintest known brown dwarf.

\section{Spitzer Detection of a Faint Companion}

\begin{figure}
\epsscale{1.15}
\plotone{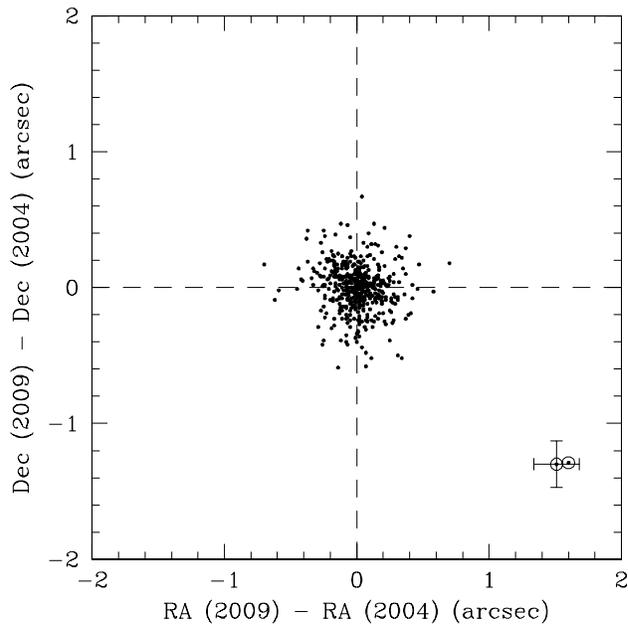}
\caption{
Differences in coordinates of sources near WD~0806-661 between two epochs
of IRAC images at 4.5~$\mu$m (points). 
We have circled WD~0806-661 and a new companion that shows the same motion.
We also include 1~$\sigma$ error bars for the latter.
}
\label{fig:pm}
\end{figure}

To search for substellar companions around nearby stars, we have considered
nearly all stars that have been observed with IRAC in multiple epochs and
that should show detectable motions in these data ($\gtrsim0.5\arcsec$,
see Figure~\ref{fig:pm}) based on their known proper motions.
This sample consists of $>$600 stars, brown dwarfs, and white dwarfs.
The plate scale and field of view of IRAC are $1\farcs2$~pixel$^{-1}$ and
$5\farcm2\times5\farcm2$, respectively. The camera produces images with
FWHM$=1\farcs6$-$1\farcs9$ from 3.6 to 8.0~\micron.
Among the four IRAC filters, we have analyzed the images collected at
3.6 and 4.5~\micron\ since they offer the best spatial resolution and
sensitivity to cool brown dwarfs. Initial processing of the images was
performed by the
pipeline at the Spitzer Science Center and the resulting data were combined
into the final images using R. Gutermuth's WCSmosaic IDL package \citep{gut08}.
Pipeline versions S18.7.0 and S18.12.0 were used for the data discussed
in this paper.

For each nearby star, we used the task {\it starfind} within IRAF to measure
positions for all point sources in the images from the multiple epochs.
We converted the pixel coordinates to equatorial coordinates using the
World Coordinate Systems produced by the pipeline and matched the 
source lists between the epochs. We then measured a new World Coordinate
System for the first epoch image using the pixel coordinates from the
first epoch and the equatorial coordinates from the second epoch, which
enabled more accurate relative astrometry between the two epochs.
To identify possible companions, we checked for sources with motions that
were consistent with the motions measured for the primaries from these data.
If accurate positions were unavailable for the primary because of saturation
in the IRAC data, we instead computed the motion expected for the primary based 
on its published proper motion and the elapsed time between the two epochs.

We have identified a new companion to the DQ white dwarf WD~0806-661 (L97-3),
which we refer to as WD~0806-661~B.
The primary was observed by IRAC on 2004 December 15 and 2009 August 24
through programs 2313 (M. Kuchner) and 60161 (M. Burleigh) with
total exposure times of 134 and 536~s, respectively.
The companion has a projected separation of $130\arcsec$, corresponding
to 2500~AU at the distance of the white dwarf \citep[19.2~pc,][]{sub09}.
The motion of WD~0806-661~B between the two epochs of IRAC data
agrees closely with that of the primary, and is significantly greater than
the zero motion expected for a background source. This is demonstrated in
Figure~\ref{fig:pm}, which shows the differences in equatorial coordinates
between the two epochs for all stars with photometric errors less than 0.1~mag
in the second epoch. For stars that have fluxes within
$\pm0.5$~mag of WD~0806-661~B, the standard deviations of the differences
in right ascension and declination in Figure~\ref{fig:pm}
are $0\farcs17$, whereas WD~0806-661~B exhibits a motion of $2\arcsec$.
The astrometric accuracy for the primary is higher since it is 3~mag
brighter than WD~0806-661~B.

\begin{figure}
\epsscale{0.5}
\plotone{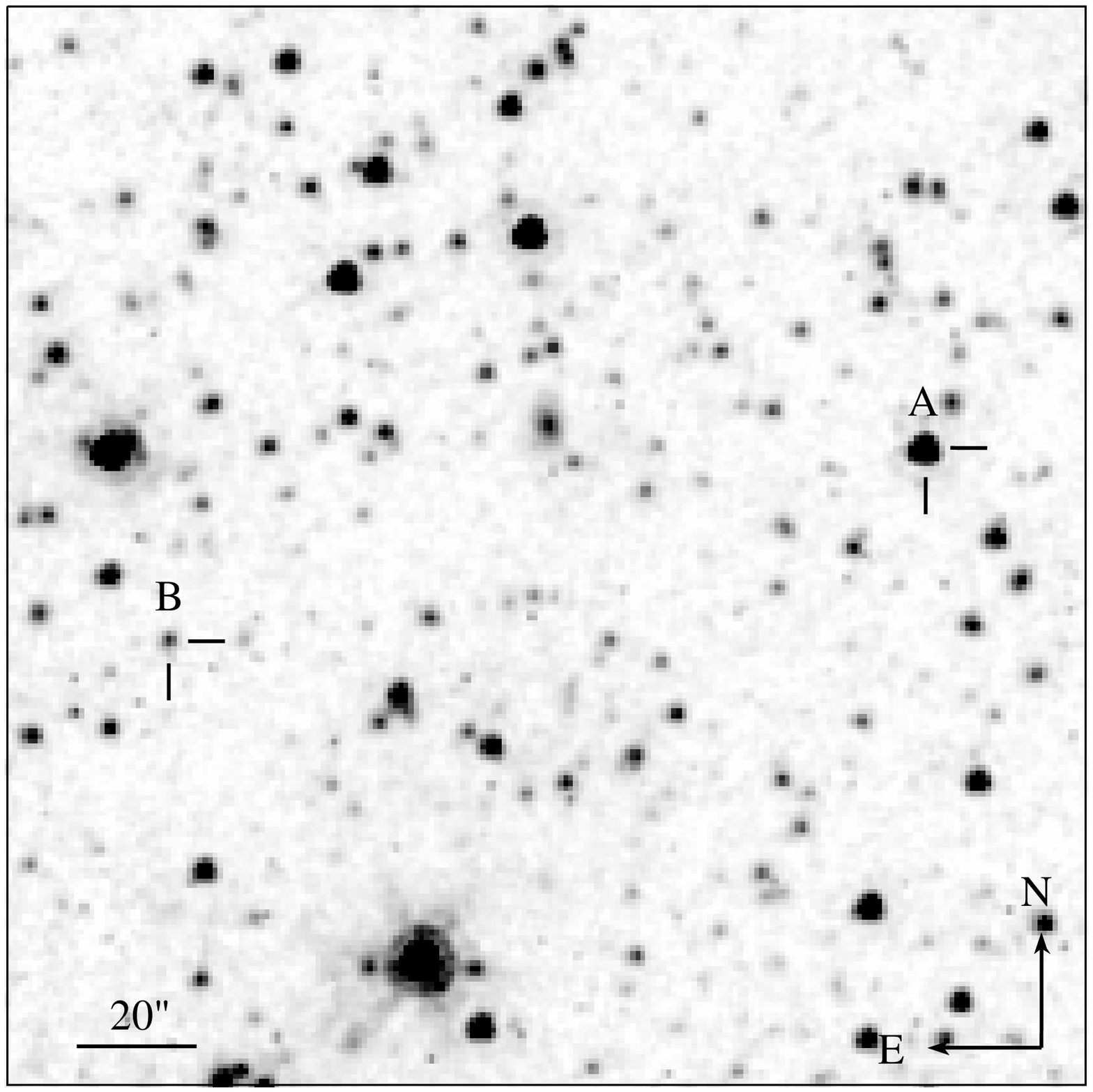}
\caption{
IRAC 4.5~\micron\ image of WD~0806-661~A and B ($3\arcmin\times3\arcmin$).
}
\label{fig:image}
\end{figure}

We measured photometry at 4.5~\micron\ for WD~0806-661~B with the methods
described by \citet{luh10tau}. 
We present the astrometry and photometry for this object as well as
other properties of the system in Table~\ref{tab:data}.
The 4.5~\micron\ image of WD~0806-661~A and B from the second epoch is
shown in Figure~\ref{fig:image}.

\section{Color Constraints}

To investigate the nature of WD~0806-661~B, we
have examined the available constraints on its colors.
It was not observed at 3.6 or 5.8~\micron\ by IRAC in either epoch.
It was encompassed by the 8.0~\micron\ images from the first epoch
but was not detected. The resulting limit of $[8.0]>15$ implies
$[4.5]-[8.0]<1.7$, which is consistent with a stellar object at
any spectral type, including a T dwarf \citep{pat06}.
We have not found any detections or useful flux limits for this object
in any publicly available images.

On 2010 December 23, we sought to obtain a spectrum of WD~0806-661~B
with the Folded-Port Infrared Echellette \citep[FIRE,][]{sim08} at 
Magellan 6.5~m Baade Telescope. However, it was not detected
in the $J$-band acquisition images, which had a total exposure time of 45~s.
By flux calibrating these images with photometry of unsaturated sources
from the Point Source Catalog of the Two-Micron All-Sky Survey
\citep{skr06}, we estimate a magnitude limit of $J>20$
for WD~0806-661~B. The combination of this limit with the IRAC measurement at
4.5~\micron\ indicates a color of $J-[4.5]>3.3$, which is
consistent with a spectral type of $\gtrsim$T8 \citep{eis10,leg10a}.
A substellar object is the only type of companion that could be as
red and faint as WD~0806-661~B.

\section{Physical Properties}
\label{sec:prop}

\begin{figure}
\epsscale{1.15}
\plotone{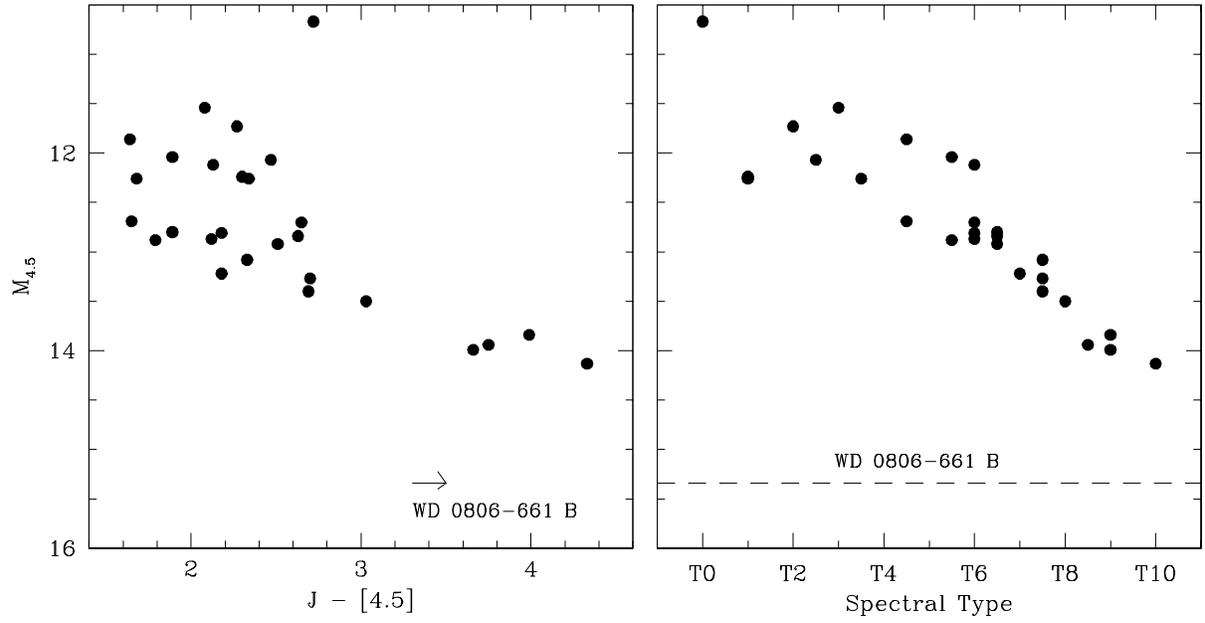}
\caption{
$M_{4.5}$ versus $J-[4.5]$ and $M_{4.5}$ versus spectral type for
WD~0806-661~B (arrow and dashed line) and
T dwarfs with measured distances and IRAC photometry
\citep[filled circles,][references therein]{luc10,leg10a}.
}
\label{fig:cmd}
\end{figure}

To estimate the physical properties of WD~0806-661~B, we begin
by assuming that it has the distance of the primary and plotting the
resulting absolute magnitude at 4.5~\micron\ versus $J-[4.5]$ and
spectral type in Figure~\ref{fig:cmd}. For comparison, we include
T dwarfs that have measured distances and IRAC photometry. These data indicate
that WD~0806-661~B should be the coolest and faintest brown dwarf
identified to date by a fairly large margin. For perspective, the sequence
of known T dwarfs spans 2.5~magnitudes while WD~0806-661~B is one magnitude
fainter than the bottom of that sequence.

As a likely companion, WD~0806-661~B should have the same age as its primary.
By combining the mass of $0.62\pm0.03$~$M_\odot$ for WD~0806-661 \citep{sub09}
with initial-final mass relations for white dwarfs \citep{kal08,wil09}, we
compute a mass of $2.1\pm0.3$~$M_\odot$ for the progenitor.
A star born with this mass should have a main sequence lifetime of
$810^{+450}_{-240}$~Myr \citep{iben89}. The sum of that lifetime and
the cooling age of 670$\pm$40~Myr from \citet{sub09} produces a 
total age of $1.5^{+0.5}_{-0.3}$~Gyr.

\begin{figure}
\epsscale{1.15}
\plotone{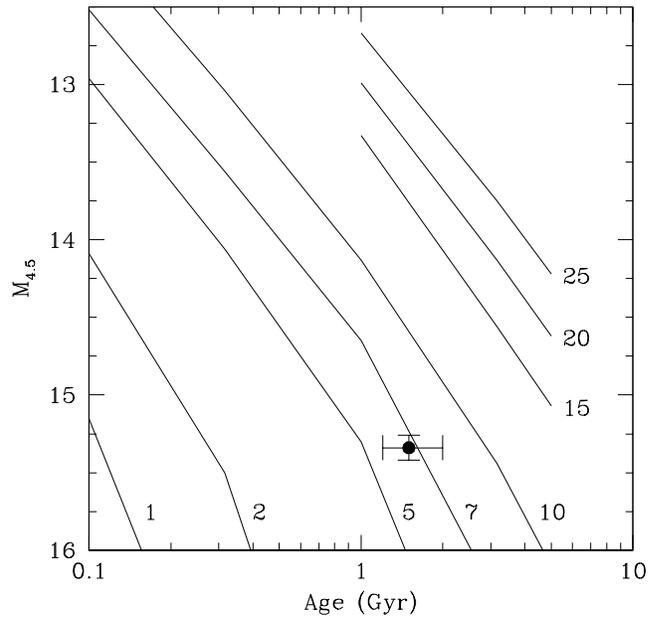}
\caption{
$M_{4.5}$ versus age for WD~0806-661~B if the distance and age of the
primary are adopted (filled circle).  For comparison, we include
the magnitudes predicted by the theoretical spectra and evolutionary models
from \citet{bur03} for constant values of mass, which are labeled
in units of $M_{\rm Jup}$ (solid lines).
}
\label{fig:mag}
\end{figure}

We can estimate the mass and temperature of WD~0806-661~B
by comparing its photometry to the fluxes predicted by theoretical
evolutionary models. In Figure~\ref{fig:mag}, we plot it
on a diagram of $M_{4.5}$ versus age under the assumption that it has
the age and distance of the primary.
\citet{bur97,bur03} used their model interiors and atmospheres to
compute effective temperatures, bolometric luminosities, and spectra
for cool brown dwarfs across a range of masses and ages.
We have convolved those synthetic spectra with the system throughput
for the 4.5~\micron\ IRAC filter. The resulting magnitudes are plotted
at constant values of mass in Figure~\ref{fig:mag}. They indicate that
WD~0806-661~B should have a mass of $\sim7$~$M_{\rm Jup}$.
The models that best match the absolute magnitude and age
of WD~0806-661~B have effective temperatures of $\sim300$~K.

\section{Discussion}

WD~0806-661~B is a strong contender for the faintest known brown dwarf
based on the available photometric and astrometric measurements.
With an estimated $T_{\rm eff}\sim300$~K, it would be significantly cooler than
the latest T dwarf currently known \citep[$T_{\rm eff}\sim500$~K][]{luc10}.
Moreover, its photosphere would be
sufficiently cool to harbor water ice clouds, making it a likely prototype for
the Y dwarf spectral class.
We also estimate a mass of $\sim7$~$M_{\rm Jup}$ for WD~0806-661~B from the
evolutionary models. Given that cloud core fragmentation appears capable
of making binary companions near this mass \citep[e.g.,][]{tod10}, it
seems likely that a companion of this kind in a very large orbit (2500~AU)
would have formed in this manner. However, the mass estimate also falls
within the range of masses measured for close-in extrasolar planets
\citep[$\lesssim15$~$M_{\rm Jup}$,][]{mar05,udr07}.  Because 
the progenitor of the primary was fairly massive ($\sim2$~$M_\odot$),
its circumstellar disk at birth could have been massive enough to give
rise to a companion at this mass. Thus, WD~0806-661~B could be a giant
planet that has been dynamically scattered to a larger orbit.

Spectroscopy and multi-band photometry are necessary to 
verify the substellar nature of WD~0806-661~B and to better
estimate its physical properties.
If confirmed as the coolest known brown dwarf, it will
represent a valuable laboratory for studying atmospheres in a new temperature
regime, and its colors will help to guide searches for the coldest brown
dwarfs with facilities like the Wide-field Infrared Survey Explorer
and the James Webb Space Telescope.

\acknowledgements

We thank Robin Ciardullo for advice regarding age estimates for white dwarfs,
telescope operator Mauricio Martinez at Magellan for his assistance with
the FIRE observations, and Nigel Hambly for his helpful referee report.
We acknowledge support from grant AST-0544588
from the National Science Foundation (K. L., J. B.) and the Chris and Warren 
Hellman Fellowship Program (A. B.). The Center for Exoplanets and Habitable
Worlds is supported by the Pennsylvania State University, the Eberly College
of Science, and the Pennsylvania Space Grant Consortium.

\clearpage

\begin{deluxetable}{ll}
\tabletypesize{\scriptsize}
\tablewidth{0pt}
\tablecaption{Properties of WD 0806-661~A and B\label{tab:data}}
\tablehead{
\colhead{Parameter} & \colhead{Value}}
\startdata
\cutinhead{WD 0806-661 A\tablenotemark{a}}
Distance & 19.2$\pm$0.6 pc \\
$\mu_{\alpha}$ & +340.3$\pm1.8$~mas~yr$^{-1}$ \\
$\mu_{\delta}$ & $-$289.6$\pm1.8$~mas~yr$^{-1}$ \\
Mass & $0.62\pm0.03$~$M_\odot$ \\
Progenitor Mass & $2.1\pm0.3$~$M_\odot$ \\
Age & $1.5^{+0.5}_{-0.3}$~Gyr \\
\cutinhead{WD 0806-661 B\tablenotemark{b}}
Separation & 130.2$\pm$0.2$\arcsec$ (2500 AU) \\
Position Angle & 104.2$\pm0.2\arcdeg$ \\
$J$ & $>$20~mag \\
$[4.5]$ & 16.75$\pm$0.05~mag \\
Est.\ $T_{\rm eff}$ & $\sim300$~K \\
Est.\ Mass & $\sim7$~$M_{\rm Jup}$ \\
\enddata
\tablenotetext{a}{Distance, $\mu$, and mass of the primary are from
\citet{sub09}. The progenitor mass and total age are estimated in
Section~\ref{sec:prop}.}
\tablenotetext{b}{This work.}
\end{deluxetable}

\begin{thebibliography}{}

\bibitem[Allard et al.(1996)]{all96}
Allard, F., Hauschildt, P. H., Baraffe, I., \& Chabrier, G. 1996, \apj, 465,
L123


\bibitem[Biller et al.(2006)]{bil06}
Biller, B. A., Kasper, M., Close, L. M., Brandner, W., \& Kellner, S. 2006,
\apj, 641, L141

\bibitem[Burgasser et al.(2006)]{bur06}
Burgasser, A. J., Geballe, T. R., Leggett, S. K., Kirkpatrick, J. D., \&
Golimowski, D. A. 2006, \apj, 637, 1067

\bibitem[Burgasser et al.(2005)]{bur05}
Burgasser, A. J., Kirkpatrick, J. D., \& Lowrance, P. J. 2005, \aj, 129, 2849

\bibitem[Burgasser et al.(2000)]{bur00}
Burgasser, A. J., et al. 2000, \apj, 531, L57


\bibitem[Burgasser et al.(2010)]{bur10}
Burgasser, A. J., et al. 2010, \apj, 725, 1405

\bibitem[Burningham et al.(2008)]{burn08}
Burningham, B., et al. 2008, \mnras, 391, 320

\bibitem[Burningham et al.(2009)]{burn09}
Burningham, B., et al. 2009, \mnras, 395, 1237

\bibitem[Burrows et al.(2003)]{bur03}
Burrows, A., Sudarsky, D., \& Lunine, J. I. 2003, \apj, 596, 587

\bibitem[Burrows et al.(1997)]{bur97}
Burrows, A., et al. 1997, \apj, 491, 856

\bibitem[Day-Jones et al.(2011)]{day11}
Day-Jones, A. C., et al. 2011, \mnras, 410, 705

\bibitem[Delorme et al.(2008)]{del08}
Delorme, P., et al. 2008, \aap, 482, 961

\bibitem[Eisenhardt et al.(2010)]{eis10}
Eisenhardt, P. R. M., et al. 2010, \aj, 139, 2455

\bibitem[Fazio et al.(2004)]{faz04}
Fazio, G. G., et al. 2004, \apjs, 154, 10

\bibitem[Geballe et al.(1996)]{geb96}
Geballe, T. R., Kulkarni, S. R., Woodward, C. E., \& Sloan, G. C. 1996, \apj,
467, L101

\bibitem[Goldman et al.(2010)]{gol10}
Goldman, B., Marsat, S., Henning, T., Clemens, C., \& Greiner, J. 2010, \mnras,
405, 1140

\bibitem[Gutermuth et al.(2008)]{gut08}
Gutermuth, R. A., et al. 2008, \apj, 674, 336

\bibitem[Iben \& Laughlin(1989)]{iben89}
Iben, I., Jr., \& Laughlin, G. 1989, \apj, 341, 312

\bibitem[Kalirai et al.(2008)]{kal08}
Kalirai, J. S., Hansen, B. M. S., Kelson, D. D., Reitzel, D. B.,
Rich, R. M., \& Richer, H. B. 2008, \apj, 676, 594

\bibitem[Kirkpatrick et al.(2005)]{kir05}
Kirkpatrick, J. D. 2005, \araa, 43, 195

\bibitem[Kirkpatrick et al.(1999)]{kir99}
Kirkpatrick, J. D. et al. 1999, \apj, 519, 802

\bibitem[Leggett et al.(2010)]{leg10a}
Leggett, S. K., et al. 2010, \apj, 710, 1627



\bibitem[Lucas et al.(2010)]{luc10}
Lucas, P. W., et al. 2010, \mnras, 408, L56

\bibitem[Luhman et al.(2010)]{luh10tau}
Luhman, K. L., Allen, P. R., Espaillat, C., Hartmann, L., \& Calvet, N.
2010, \apjs, 186, 111


\bibitem[Luhman et al.(2007)]{luh07}
Luhman, K. L., et al. 2007, \apj, 654, 570


\bibitem[Marcy et al.(2005)]{mar05}
Marcy, G., et al. 2005, Progress of Theoretical Physics Supplement, 158, 24

\bibitem[Marley et al.(1996)]{mar96}
Marley, M. S., et al. 1996, Science, 272, 1919

\bibitem[McCaughrean et al.(2004)]{mcc04}
McCaughrean, M. J., et al. 2004, \aap, 413, 1029

\bibitem[Metchev et al.(2008)]{met08}
Metchev, S., Kirkpatrick, J. D., Berriman, G. B., \& Looper, D. 2008,
\apj, 676, 1281

\bibitem[Mugrauer et al.(2006)]{mug06}
Mugrauer, M., Seifahrt, A., Neuh\"{a}user, R., \& Mazeh, T. 2006, \mnras, 373,
L31

\bibitem[Nakajima et al.(1995)]{nak95}
Nakajima, T., Oppenheimer, B. R., Kulkarni, S. R., Golimowski, D. A.,
Matthews, K., \& Durrance, S. T. 1995, \nat, 378, 463

\bibitem[Oppenheimer et al.(1995)]{opp95}
Oppenheimer, B. R., Kulkarni, S. R., Nakajima, T., \& Matthews, K. 1995,
Science, 270, 1478

\bibitem[Patten et al.(2006)]{pat06}
Patten, B. M., et al. 2006, \apj, 651, 502

\bibitem[Scholz et al.(2003)]{sch03}
Scholz, R.-D., McCaughrean, M. J., Lodieu, N., \& Kuhlbrodt, B. 2003, \aap,
398, L29

\bibitem[Scholz(2010)]{sch10}
Scholz, R.-D. 2010, \aap, 515, A92

\bibitem[Simcoe et al.(2008)]{sim08}
Simcoe, R. A. et al. 2008, Proc. SPIE, 7014, 27

\bibitem[Skrutskie et al.(2006)]{skr06}
Skrutskie, M., et al. 2006, \aj, 131, 1163

\bibitem[Subasavage et al.(2009)]{sub09}
Subasavage, J. P., Jao, W.-C., Henry, T. J., Bergeron, P., Dufour, P.,
Ianna, P. A., Costa, E., \& M\'{e}ndez, R. A. 2009, \aj, 137, 4547

\bibitem[Todorov et al.(2010)]{tod10}
Todorov, K., Luhman, K. L., \& McLeod, K. K. 2010, \apj, 714, L84

\bibitem[Udry et al.(2007)]{udr07}
Udry, S., Fischer, D., \& Queloz, D. 2007, Protostars and Planets V, B.
Reipurth, D. Jewitt,and K. Keil (eds.), University of Arizona Press,
Tucson, 685


\bibitem[Werner et al.(2004)]{wer04}
Werner, M. W., et al. 2004, \apjs, 154, 1

\bibitem[Williams et al.(2009)]{wil09}
Williams, K., Bolte, M., \& Koester, D. 2009, \apj, 693, 355

\bibitem[Wilson et al.(2001)]{wil01}
Wilson, J. C., Kirkpatrick, J. D., Gizis, J. E., Skrutskie, M. F.,
Monet, D. G., \& Houck, J. R. 2001, \aj, 122, 1989

\end{thebibliography}
\end{document}